\documentclass[a4paper,11pt]{article}

\usepackage{jheppub, bm, color} 
\usepackage{amssymb,amsfonts,slashed,amsthm,amsmath,graphicx, soul}
\usepackage[caption=false]{subfig}

\usepackage{hyperref}

\usepackage{epsfig}

\newcommand{\eV}{ \ {\rm eV} }

\newcommand{\MeV}{\  {\rm MeV} }
\newcommand{\GeV}{\  {\rm GeV} }

\newcommand{\lmk}{\left(}  
\newcommand{\rmk}{\right)}
\newcommand{\lkk}{\left[}  
\newcommand{\rkk}{\right]}

\newcommand{\del}{\partial}  
\newcommand{\la}{\left\langle} 
\newcommand{\ra}{\right\rangle}

\newcommand{\bea}{\begin{align}}
\newcommand{\eea}{\end{align}}
\newcommand{\beq}{\begin{equation}}
\newcommand{\eeq}{\end{equation}}

\newcommand{\pp}{p}

\newcommand{\Mpl}{M_{\rm Pl}}
\newcommand{\abs}[1]{\left\vert {#1} \right\vert}

\newcommand{\eq}[1]{Eq.~(\ref{#1})}

\newcommand{\aex}{a_{\rm exit}}
\newcommand{\aRH}{a_{\rm RH}}
\newcommand{\aNRl}{a_{\rm NR,3}}
\newcommand{\aNRe}{a_{\rm NR,1}}
\newcommand{\aNRm}{a_{\rm NR,2}}
\newcommand{\ain}{a_{\rm in}}
\newcommand{\ac}{a_{\rm c}}
\newcommand{\astar}{a_{*}}
\newcommand{\astars}{a_{**}}
\newcommand{\anow}{a_{\rm eq}}

\newcommand{\aNR}{a_{\rm NR,RD}}
\newcommand{\aNRi}{a_{\rm NR,inf}}
\newcommand{\arela}{a_{\rm rela}}

\newcommand{\HI}{H_{I}}
\newcommand{\Hnow}{H_{\rm eq}}

\newcommand{\kRH}{k_{\rm RH}}
\newcommand{\kstar}{k_{*}}
\newcommand{\kstars}{k_{**}}

\newcommand{\g}{\tilde{g}}
\newcommand{\m}{m_{\gamma',0}}
\newcommand{\mg}{m_{\gamma'}}
\newcommand{\mginf}{m_{\gamma',{\rm inf}}}

\newcommand{\Aex}{A_{L,0}}

\begin{document}

\begin{flushright}
TU-1151
\end{flushright}

\title{
Gravitational production of dark photon dark matter with mass generated by the Higgs mechanism
}

\author{Takanori Sato$^1$, 
Fuminobu Takahashi$^{1}$, 
Masaki~Yamada$^{1,2}$
\\*[20pt]
$^1${\it \normalsize
Department of Physics, Tohoku University, Sendai, Miyagi 980-8578, Japan}  \\*[5pt]
$^2${\it \normalsize 
FRIS, Tohoku University, Sendai, Miyagi 980-8578, Japan} 
}
\emailAdd{takanori.satou.p4@dc.tohoku.ac.jp}
\emailAdd{fumi@tohoku.ac.jp}
\emailAdd{m.yamada@tohoku.ac.jp}

\abstract{
We study the gravitational production of dark photon dark matter during inflation, when dark photons acquire mass by the Higgs mechanism. In the previous study, it was assumed that the dark photon has a  St\"uckelberg mass, or a mass generated by the Higgs mechanism with a sufficiently heavy Higgs boson. In this paper we consider a case in which the Higgs boson is not fully decoupled; the Higgs field changes its vacuum expectation value after inflation. Then, the dark photon mass also changes with time after inflation, and the time evolution of the longitudinal mode is different from the case with a St\"{u}ckelberg mass. Consequently, the spectrum of the dark photon energy density can have two peaks at an intermediate scale and a small scale. We show that the dark photon can explain the dark matter if its current mass is larger than $6 \, \mu {\rm eV} \times (H_I / 10^{14} \GeV)^{-4}$ and smaller than $0.8 \GeV \times (H_I / 10^{14} \GeV)^{-3/2}$, with $H_I$ being the Hubble parameter during inflation. A higher mass is required if one considers a larger gauge coupling constant. The result for the  St\"uckelberg mass can be reproduced in the limit of a small gauge coupling constant. We also comment on the constraints set by various conjectures in quantum gravity theory.
}

\maketitle

\section{Introduction}

The existence of dark matter (DM) has been confirmed by a plethora of astrophysical and cosmological observations. The characteristics of DM such as  being cold, electrically neutral, and stable have also been reported. However, fundamental questions remain: what is the DM and how is it produced in the early universe? 
The allowed mass of the DM ranges from $10^{-22} \eV$ 
to astronomical mass scales,
and even the spin of the DM remains undetermined. 
The DM may be produced thermally or non-thermally. 
Weakly-interacting DM is 
one of the plausible DM candidates, but
it is still undiscovered and raises questions about its existence and sets tight restrictions on new physics at the electroweak scale.
Therefore, it is important to investigate other candidates for DM.

Among the DM models, a vector boson or dark photon has recently attracted much attention. It is a gauge boson of a dark  U(1)$'$ gauge symmetry, which generically has a kinetic mixing with the U(1)$_Y$ gauge boson~\cite{Holdom:1985ag,Gherghetta:2019coi}.
 If the mass of the dark photon is smaller than twice the electron mass or if the kinetic mixing is sufficiently small, it can be stable on cosmological time scales.
Numerous experiments are currently underway to search for dark photon DM directly or indirectly through the kinetic mixing~\cite{Antypas:2022asj}.

A very light dark photon should be produced non-thermally so that it is sufficiently cold well before the matter-radiation equality. 
A relatively heavy dark photon may also have to be produced non-thermally because the kinetic mixing should be extremely small to ensure a sufficiently long lifetime and it is difficult to thermalize dark photons. There are various non-thermal production mechanisms using e.g. the
dynamics of cosmic strings~\cite{Long:2019lwl}, 
phase transition~\cite{Nakayama:2021avl}, 
preheating and decay of axion-like particles~\cite{Agrawal:2018vin,Dror:2018pdh,Co:2018lka,Bastero-Gil:2018uel}, 
and gravitational production during and after inflation~\cite{Graham:2015rva,Ema:2019yrd,Ahmed:2020fhc,Nakayama:2020ikz,Kolb:2020fwh}. 
The misalignment mechanism~\cite{Nelson:2011sf,Arias:2012az} requires a non-minimal coupling to gravity, which however does not work because of the ghost instability~\cite{Nakayama:2019rhg}, and it is not a viable option for realistic models~\cite{Nakayama:2020rka}.

The gravitational production of dark photon DM during inflation was first investigated in Ref.~\cite{Graham:2015rva} for the case in which the dark photon has a St\"{u}ckelberg mass. Their results also hold when the mass is generated by the Higgs mechanism if the Higgs boson (radial mode) is so heavy that it is effectively decoupled during and after inflation.
They found that the  longitudinal mode of dark photon 
is efficiently generated from quantum fluctuations during inflation. 
The spectrum of dark photon energy density has a peak at an intermediate scale, which gives the dominant contribution to the abundance. Thus produced dark photon can explain the observed DM abundance with the dark photon mass satisfying $6 \, \mu {\rm eV} \times (H_I / 10^{14} \GeV)^{-4}$, where $H_I$ is the Hubble parameter during inflation. The large hierarchy between the inflationary scale and the mass of the dark photon suggests that the gauge coupling is very small, given the
bound on the radial mode mass. Assuming that the dark photon explains all DM, the gauge coupling must be smaller than $\sim 10^{-28} (H_I/10^{14} \GeV)^{-5}$.
Such a small gauge coupling constant is known to limit the viable dark photon mass  based on various conjectures in quantum gravity~\cite{Reece:2018zvv}.

We consider the gravitational production of dark photon DM for the case in which the dark photon obtains a mass via the Higgs mechanism and the Higgs boson is not fully decoupled during and after inflation. The dynamics of the Higgs boson could
complicate the evolution of the system in various ways. If the Higgs boson is much lighter than $H_I$, it may be displaced from the origin during inflation.
After inflation it starts to oscillate about the origin  with a large initial amplitude, and the dark U(1)$'$ symmetry may be restored after inflation. 
Once the symmetry is restored, the subsequent evolution of dark photon would be
completely different from that considered in the gravitaional production, and it would be similar to the scenario using the cosmic strings~\cite{Long:2019lwl}. On the other hand, it is more natural to consider that the Higgs boson acquires a mass of order the Hubble parameter in the early Universe. 
Such an effective mass is often referred to as the Hubble-induced mass and arises naturally from interactions with the inflaton and/or non-minimum coupling with gravity~\cite{Dine:1995uk}. If the Hubble-induced mass is negative, the Higgs field will develop
a large vacuum expectation value (VEV) during inflation. After inflation,
the potential minimum of the Higgs field changes with time, and
it eventually reaches the present vacuum when the Hubble parameter decreases to the bare mass of the Higgs field. 
Such dynamics of the Higgs field leads to a time-dependent mass for the dark photon. 
As a result, the dark photon evolves differently from the case with a 
St\"{u}ckelberg mass.

In this paper, we focus on the consequence of time-dependent dark photon mass for the gravitational production of dark photon. To this end, we will consider a simple potential for the Higgs field, and parametrize the Higgs VEV as a power-law function of time in our analysis, without giving the detailed origin of such potential of the Higgs sector. Later we will comment on a instability of the Higgs dynamics for some choice of the potential, and a possible solution to avoid the problem.
By solving the equation of motion for the longitudinal mode to calculate its present spectrum, we will determine those parameters through which we could explain DM abundance by gravitationally produced dark photons. 
As expected, our results are reduced to the case of Ref.~\cite{Graham:2015rva} in the limit of the small gauge coupling constant, and our scenario provides a larger parameter space for a fixed gauge coupling constant. Our result can be applied to any model in which the Higgs VEV changes after inflation with some power of time variable and becomes constant at a certain time.

The rest of this paper is organized as follows. 
In Sec.~\ref{sec:evolution}, we analytically solve the evolution of dark photons in each regime of interest and calculate its spectrum. 
In Sec.~\ref{sec:DM}, we determine the parameter region in which we can explain the DM abundance. We also clarify the relation to the case with the St\"{u}ckelberg mechanism.
Section~\ref{sec:conclusion} is devoted to the discussion and conclusions. In the Appendix we discuss the case of a quartic potential for the Higgs boson, and comment on the instability of the dynamics.

\section{Evolution of dark photon}
\label{sec:evolution}

We consider a dark photon that acquires a mass by the Higgs mechanism. The action is given by 
\begin{equation}
 S = \int \sqrt{-{\rm det} \, g } \, d^4x \lkk 
 - \frac14 g^{\mu \rho}g^{\nu \sigma} F'_{\mu \nu} F'_{\rho \sigma} 
 - \abs{(\del_\mu - i g A'_\mu) \Phi}^2
 - V(\Phi) 
 \rkk\,, 
\end{equation}
where $F'_{\mu \nu}$, $A'_\mu$, $g$, and $\Phi$ denote the field strength of U(1)$'$, dark photon, U(1)$'$ gauge coupling, and the Higgs boson, respectively, while $V(\Phi)$ represents the Higgs boson potential. 
Throughout this paper, we consider the instantaneous reheating, where inflation is followed by a radiation-dominated epoch. 

As discussed in the introduction, 
we consider the case in which the VEV of the Higgs field changes after inflation while the U(1)$'$ gauge symmetry is spontaneously broken throughout the history of the Universe. 
For example, 
one may consider a wine-bottle-type potential for $V(\Phi)$ 
with a negative Hubble-induced mass term: 
\begin{equation}
\label{Vphi}
 V(\Phi) = - c_H H^2 \abs{\Phi}^2 
 - m_\Phi^2 \abs{\Phi}^2 + \frac{1}{\Lambda^{2n-4}} \abs{\Phi}^{2n}\,, 
\end{equation}
where $c_H$ is a positive constant, $m_\Phi$ is the (bare) mass, $\Lambda$ is the cut-off scale of the interaction, and $n$ ($\ge 2$) is an integer. 
The most natural value of $c_H$ is of order unity.
The Hubble-induced mass naturally arises from  Planck-suppressed couplings with the inflaton and/or non-minimal coupling to gravity; e.g. it generically appears via supergravity effects in supersymmetric models~\cite{Dine:1995uk}.%
\footnote{
During the radiation dominated era, 
the Hubble-induced mass term could be suppressed by the approximate conformal invariance of radiation. 
We assume that $c_H \gtrsim \mathcal{O}(1)$, after considering the suppression factor~\cite{Kawasaki:2012rs}.
}
The higher-dimensional term with suppressed lower-dimensional terms can also be expected in supersymmetric flat directions. 
In this case, the potential minimum of $\Phi$ is given by 
\begin{equation}
 \Phi_{\rm min}(t) \simeq 
 \left\{
 \begin{aligned}
 &\lmk \frac{c_H }{n } \Lambda^{2n-4} H^2 \rmk^{1/(2n-2)}, 
 &\quad \text{for} \ \sqrt{c_H} H \gg m_\Phi \,, 
 \\
&\lmk \frac{1}{n } \Lambda^{2n-4} m_\Phi^2 \rmk^{1/(2n-2)}, 
 &\quad \text{for} \ \sqrt{c_H} H \ll m_\Phi \,. 
\end{aligned}
\right.
 \label{vev}
\end{equation}
We consider the case with $H_I \gg m_\Phi$. 
During inflation, since the Higgs mass is of order the Hubble parameter, the Higgs stays at the potential minimum, $\la \Phi \ra = \Phi_{\rm min}$, and U(1)$'$ symmetry is kept broken. 
One can expect that, for $c_H \gtrsim O(1)$, the Higgs field follows its potential minimum after inflation though its dynamics could be non-trivial for $n=2$; 
one has to take into account the growth of fluctuations of Higgs field and the thermal potential from the dark photon. We will discuss this case in Appendix~\ref{sec:appendix}.
In this paper, we focus on the effect of the time-dependent dark photon mass
for the gravitational production, and to this end, we 
do not specify the UV origin of the Higgs sector.  Instead, we will simply
parametrize the evolution of the Higgs VEV and dark photon mass as a power-law
function of the cosmic time.

After inflation, the Higgs VEV as well as the dark photon mass decrease with time, and at a certain point,  the Higgs field reaches the present potential minimum.
Let us denote this timing by $t_c$. It must be well before the matter-radiation equality. 
Consequently, the dark photon mass is time-dependent and given by 
\begin{align}
 m_{\gamma'}(t) &
 = \sqrt{2}\, g \la \Phi(t) \ra, 
 \\
&=
\begin{cases}
\displaystyle
\mginf, &\quad \text{for}\   t <t_{\rm RH}
\\
\displaystyle
\g \HI
\lmk \frac{t}{t_{\rm RH}} \rmk^{-\pp}\,, &\quad \text{for}\  t_{\rm RH} < t < t_c
\\
\displaystyle
\m &\quad \text{for}\ t > t_c, 
\end{cases}
\label{eq:mass}
\end{align}
where $\mginf$ ($\m$) is the dark photon mass during inflation (at present), 
$\pp$ and $\g$ are constants, $t_{\rm RH}$ is the time at the reheating, and $\HI$ is the Hubble parameter during inflation. 
If one considers the Higgs potential of \eq{Vphi}, $\pp = 1/(n-1)$ and $\g = \sqrt{2}\, g (c_H /n )^{1/(2n-2)}(\Lambda/\HI)^{(n-2)/(n-1)}$. 
We assume $\mginf \ll \HI$ to gravitationaly produce the dark photon during inflation. 
We expect $\la \Phi(t) \ra \gtrsim \HI$ 
from the unitarity of the Higgs coupling 
because the Higgs mass should be larger than or comparable to $\HI$ during inflation. 
This implies $g \lesssim \g \ll 1$. 
The variables should satisfy the following relations so that the dark photon mass changes continuously: 
\begin{align}
 &\m = \g \HI
\lmk \HI t_c \rmk^{-\pp}\,, 
\\
&\mginf = \g \HI\,,
\end{align}
where we use $t_{\rm RH} \sim H_{I}^{-1}$.  
The free parameters are thus three of $\mginf$, $\m$, $\g$, $\pp$, and $t_c$. 
We use $\mginf$, $\m$, and $\pp$ to calculate the spectrum but rewrite $\mginf$ in temrs of $\g$ to demonstrate the parameter space of interest. 
In the rest of this section, we solve the equation of motion for the dark photon that is gravitationally produced during inflation.

\subsection{Action and equation of motion}

The action for the dark photon can be decomposed into transverse and longitudinal modes such as 
\begin{equation}
\label{transverse}
S_{\rm{T}}=\int\frac{a^3 \, d^3k \, dt}{(2\pi)^3}\frac{1}{2a^2} \lkk 
\abs{\partial_{t}A^i_{T}}^2-\lmk {k^2}{a^2}+m_{\gamma'}^2\rmk \abs{A^i_{T}}^2\rkk \,,
\end{equation}
and
\begin{equation}
\label{longitudinal_action}
S_{\rm{L}}=\int\frac{a^3\, d^3k\, dt}{(2\pi)^3}\frac{1}{2a^2}\lkk \frac{a^2m_{\gamma'}^2}{k^2+a^2 m_{\gamma'}^2}\abs{\partial_{t}A_{L}}^2-m_{\gamma'}^2\abs{A_{L}}^2\rkk \,,
\end{equation}
where we take a Fourier transformation and define 
\begin{align}
&\sum_i k^i A^i=kA_{L}\\
&\sum_i k^i A^i_{T}=0 \,,
\end{align}
with $k = \abs{k^i}$. 
As discussed in Ref.~\cite{Graham:2015rva}, although the transverse mode does not grow owing to the approximate conformal invariance for a small $m_{\gamma'}$, the longitudinal mode can grow efficiently. Hence, we focus on the latter dynamics.

In the relativistic limit, $m_{\gamma'} \ll k/a$, $S_L$ can be simplified by defining 
\begin{equation}
 \label{rescale}
\pi (k^i,t)\equiv \frac{\mg}{k}A_{L}(k^i,t) \,,
\end{equation}
such as 
\begin{equation}
\label{massless scalar}
S_{\rm{L}} \approx \int \frac{a^3\, d^3k\, dt}{(2\pi)^3}\lkk \frac{1}{2}|\partial_{t}\pi|^2-\frac{k^2}{2a^2}|\pi|^2\rkk \,. 
\end{equation}
This implies that the vacuum state of $\pi$ during inflation is given by the Bunch-Davis vacuum expressed as 
\begin{equation}
\pi(k^i,t)=\pi_{0}(k^i)\lkk 1-\frac{ik}{aH_{I}} \rkk \rm{exp}\lmk \frac{\textit{ik}}{{\textit{aH}_{\textit{I}}}} \rmk \,. 
\end{equation}
Here, $\pi_0$ satisfies 
\begin{equation}
\mathcal{P}_{\pi_0}(k)=\lmk\frac{H_{I}}{2\pi}\rmk^2 \,, 
\end{equation}
where the power spectrum is defined by 
\begin{align}
\langle \pi_0^{*}(\vec{k}, t) \pi_0(\vec{k}^{'}, t)\rangle 
\equiv(2\pi)^3\delta^3(\vec{k}-\vec{k'})\frac{2 \pi^2}{k^3}\mathcal{P}_{\pi_0}(k,t) \,. 
\end{align}
Considering $A_L$ again, we obtain 
\begin{align}
\label{eom for AL}
A_{L}(\vec{k},t)
&=\Aex(\vec{k})\lkk1-\frac{ik}{aH_{I}} \rm{exp}\lmk\frac{\textit{ik}}{{\textit{aH}_{\textit{I}}}}\rmk \rkk \,,
\end{align}
with
\begin{equation}
\mathcal{P}_{\Aex}(k)=\lmk\frac{kH_{I}}{2\pi \mginf}\rmk^2 \,. 
\end{equation}
This sets the initial condition of $A_L$ during inflation. Here we note that the dark photon mass during inflation appears in the denominator, which will be important when comparing the abundance with the case of a St\"ueckelberg mass.

The energy density of $A_L$ can be calculated from 
\begin{equation}
\label{energy density spectrum}
 \rho_{A_L}(t)=\int d \ln k \, \frac{1}{2a^2}\lkk\frac{a^2\mg^2}{k^2+a^2\mg^2}\mathcal{P}_{\partial A_{L}}(k,t)+\mg^2\mathcal{P}_{A_{L}}(k,t)\rkk \,.
\end{equation}
At a later epoch, the dark photon mass becomes
constant when the Hubble parameter becomes sufficiently small. 
Then, the energy density of the dark photon can be calculated from 
\begin{equation}
\label{energy density spectrum2}
 \rho_{\gamma'}(t) \simeq \int d \ln k \, \frac{\m^2}{a^2} \mathcal{P}_{A_{L}}(k,t) \,.
\end{equation}
Because the longitudinal mode evolves with time, the power spectrum depends on time as defined by 
\begin{equation}
\label{spectrum1}
\mathcal{P}_{A_{L}}(k,t)= \lmk\frac{A_{L}(t)}{\Aex}\rmk^2 \mathcal{P}_{\Aex}(k)= \lmk\frac{A_{L}(t)}{\Aex}\rmk^2 \lmk\frac{kH_{I}}{2\pi \mginf}\rmk^2 \,. 
\end{equation}
Hence, we need to calculate the present value of $A_L(t)$ by solving its equation of motion.

\subsection{Analytic solutions}

From \eq{longitudinal_action}, 
the equation of motion for the longitudinal mode can be expressed as
\begin{equation}
 \lmk \partial_{t}^2+\frac{\lmk 3H+2 (\partial_t m_{\gamma'})/\mg \rmk k^2+a^2\mg^2H}{k^2+a^2\mg^2}\partial_{t}+\frac{k^2}{a^2}+\mg^2\rmk A_{L}=0 \,,
 \end{equation}
where the time-dependent mass $m_{\gamma'}(t)$ provides an additional (anti-) friction term. 
This leads to a different dynamics for the dark photon, compared with the case involving a constant mass term, such as the case in the St\"{u}ckelberg mechanism.

\begin{figure}[t]
\begin{center}
\includegraphics[clip, width=13cm]{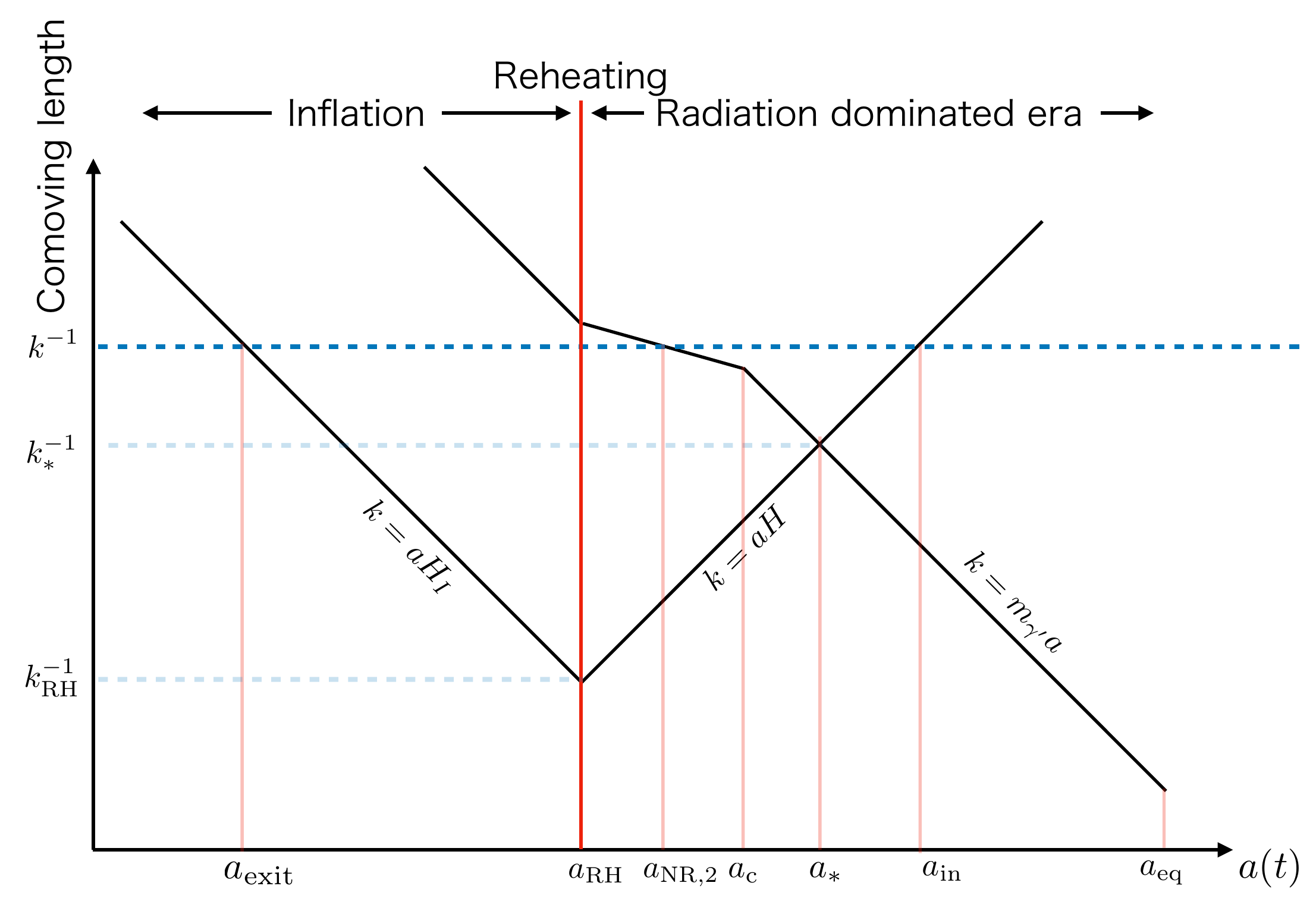}
\end{center}
\vspace{0.3cm}
\begin{center}
\includegraphics[clip, width=13cm]{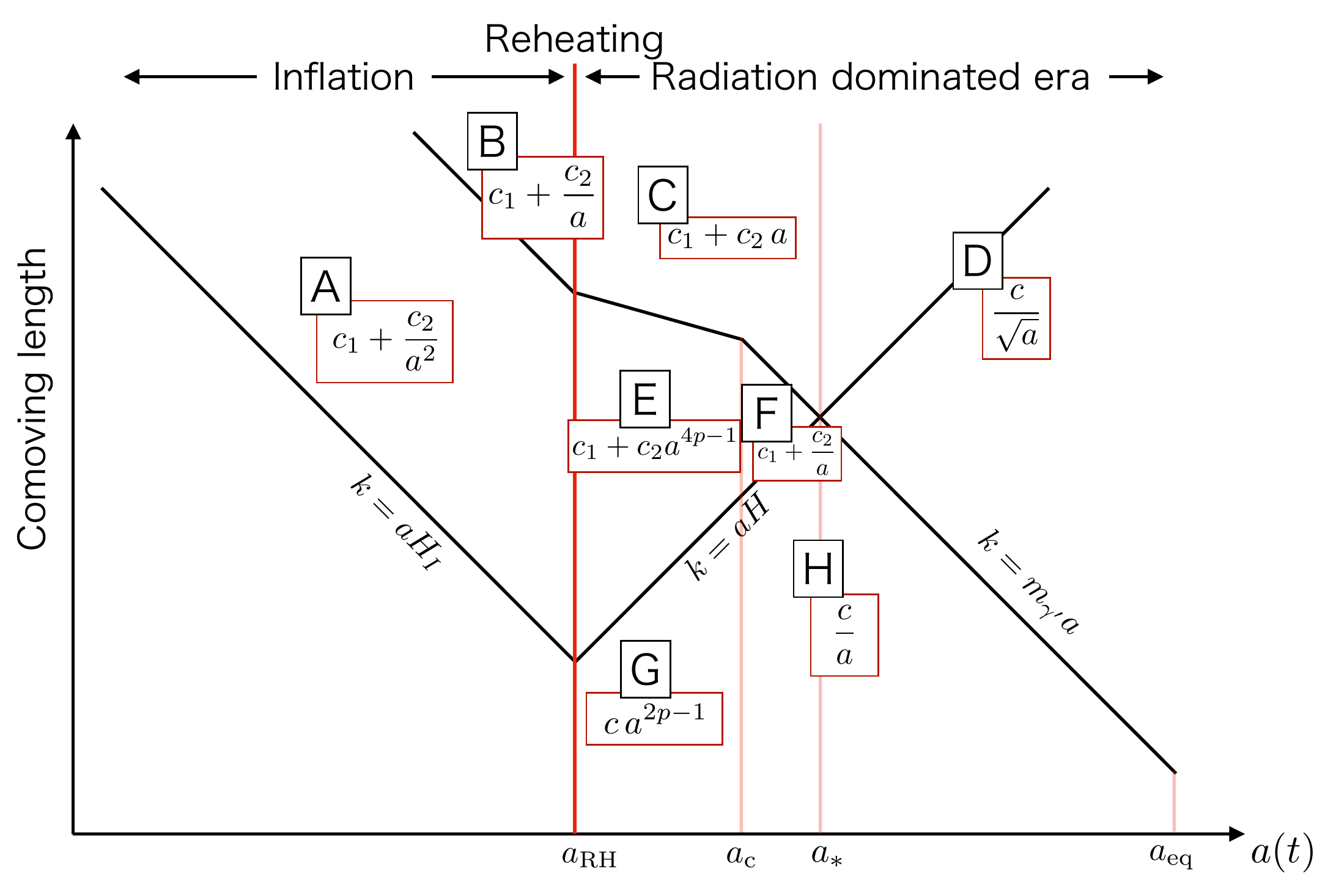}
\end{center}
\caption{ Schematic picture of comoving length evolution for Case 1. 
The boxes in the lower panel represent the evolution of the longitudinal mode of dark photon $A_L(a)$. The constants $c_i$ and $c$ should be understood as different values in different regimes and determined by connecting the solutions on the boundaries. 
\label{fig:1}
}
\end{figure}

We follow the discussion in Ref.~\cite{Graham:2015rva} 
and divide the evolution of dark photon into the regimes in which we can analytically solve the equation of motion. 
The equation of motion can be simplified by taking super-horizon/sub-horizon and relativistic/non-relativistic limits. 
The threshold of the former is simply given by $k = a(t) H(t)$. 
A mode with the comoving wavenumber $k$ exits the horizon during inflation at $a(t) = \aex \equiv (k/\kRH)\aRH$, and later enters into the horizon after inflation at $a(t) = \ain \equiv (\kRH/k)\aRH$. 
Here, we define $\aRH$ as the scale factor at the reheating and $\kRH$ as the comoving wavenumber which enters into the horizon at the reheating. 
The threshold of the relativistic/non-relativistic limit is complicated owing to the time-dependence of the dark photon mass. 
A mode with a relatively small $k$ 
can become non-relativistic during inflation at $a(t) = \aNRe \equiv \aRH (k/\kRH) (H_I/\mginf)$ because its physical wavenumber $k/a(t)$ decreases in time while the dark photon mass is constant. 
After inflation, the dark photon mass decreases as $\propto a(t)^{-2\pp}$, which is slower than $a(t)^{-1}$ for $\pp < 1/2$, such that a mode at an intermediate scale can become non-relativistic at $a(t) = \aNRm \equiv \aRH (k/\kRH)^{1/(1-2p)} (H_I/\mginf)^{1/(1-2p)}$. 
Here and hereafter, we focus on the case with $\pp < 1/2$. The case with $\pp = 1$ is explained in Appendix~\ref{sec:appendix}, where we find that the resulting dark photon abundance has the same parameter dependence with the case of $\pp < 1/2$. 
At $a(t) = \ac \equiv \aRH (\mginf/\m)^{1/(2p)}$,
the Higgs reaches the low-energy minimum and the dark photon mass becomes constant. 
Subsequently, 
a mode with a relatively large $k$ can become non-relativistic at $a = \aNRl \equiv \aRH (k/\kRH)(\HI/\m)$. 
This dependence is summarized in the upper panel of Fig.~\ref{fig:1}. 
The vertical red line represents the reheating epoch at $a(t) = \aRH$.

The scale factors defined above are summarized as 
\begin{align}
\label{eq:scale1}
 &\frac{\aex}{\aRH} = \lmk \frac{k}{\kRH} \rmk,
\qquad 
 \frac{\aNRe}{\aRH} = \lmk \frac{k}{\kRH} \rmk 
 \lmk \frac{H_I}{\mginf} \rmk, 
\\
\label{eq:scale2}
 &\frac{\aNRm}{\aRH} = 
 \lmk \frac{k}{\kRH} \rmk^{1/(1-2p)} 
 \lmk \frac{H_I}{\mginf} \rmk^{1/(1-2p)}, 
\qquad
 \frac{\aNRl}{\aRH} = \frac{k}{\kRH} \lmk \frac{\HI}{\m} \rmk,
\\
\label{eq:scale3}
 &
 \frac{\ain}{\aRH} = \lmk \frac{\kRH}{k} \rmk,
\qquad 
 \frac{\ac}{\aRH} = 
 \lmk \frac{\mginf}{\m} \rmk^{1/(2p)},
\end{align}
For convenience, we define 
the scale factors as 
\begin{equation}
 \frac{\astar}{\aRH} = \lmk \frac{\HI}{\m} \rmk^{1/2} 
\qquad
 \frac{\astars}{\aRH} = \lmk \frac{\HI}{\mginf} \rmk^{1/(2-2p)} 
\end{equation}
and 
the comoving wavenumbers as 
\begin{equation}
 \kstar = \aRH \sqrt{\m \HI},
\qquad
 \kstars = 
 \aRH \HI \lmk \frac{\mginf}{H_I} \rmk^{1/(2-2p)}, 
\qquad
 \kRH = \aRH \HI \,,
\end{equation}
which satisfy 
\begin{equation}
 \frac{\kstar}{\kRH} = \lmk \frac{\m}{\HI} \rmk^{1/2} \,,
 \qquad
  \frac{\kstars}{\kRH} = \lmk \frac{\mginf}{\HI} \rmk^{1/(2-2p)} \,
\end{equation}
The wavenumber $\kRH / \aRH$ represents the mode that is as large as the inverse of the horizon length at the end of inflation. 
The wavenumber $\kstar$ ($\kstars$) is the mode that enters into the horizon at the time when it becomes non-relativistic 
and 
the corresponding scale factor is given by $a_*$ ($a_{**})$. 
We define $k_*$ and $k_{**}$ such that $a_* > a_c$ (Case 1) and $a_{**} < a_c$ (Case 2), 
where 
\begin{align}
&{\rm Case \ 1}: 
\label{eq:mginfth1}
\mginf < \mginf^{\rm (th)} \equiv \HI \lmk \frac{\m}{\HI} \rmk^{1-p}
\\
&{\rm Case \ 2}: 
\mginf > \mginf^{\rm (th)}. 
\label{eq:mginfth2}
\end{align}
We will see shortly that the shape of the spectrum is different in those regimes.

\begin{figure}[t]
\begin{center}
\includegraphics[clip, width=14cm]{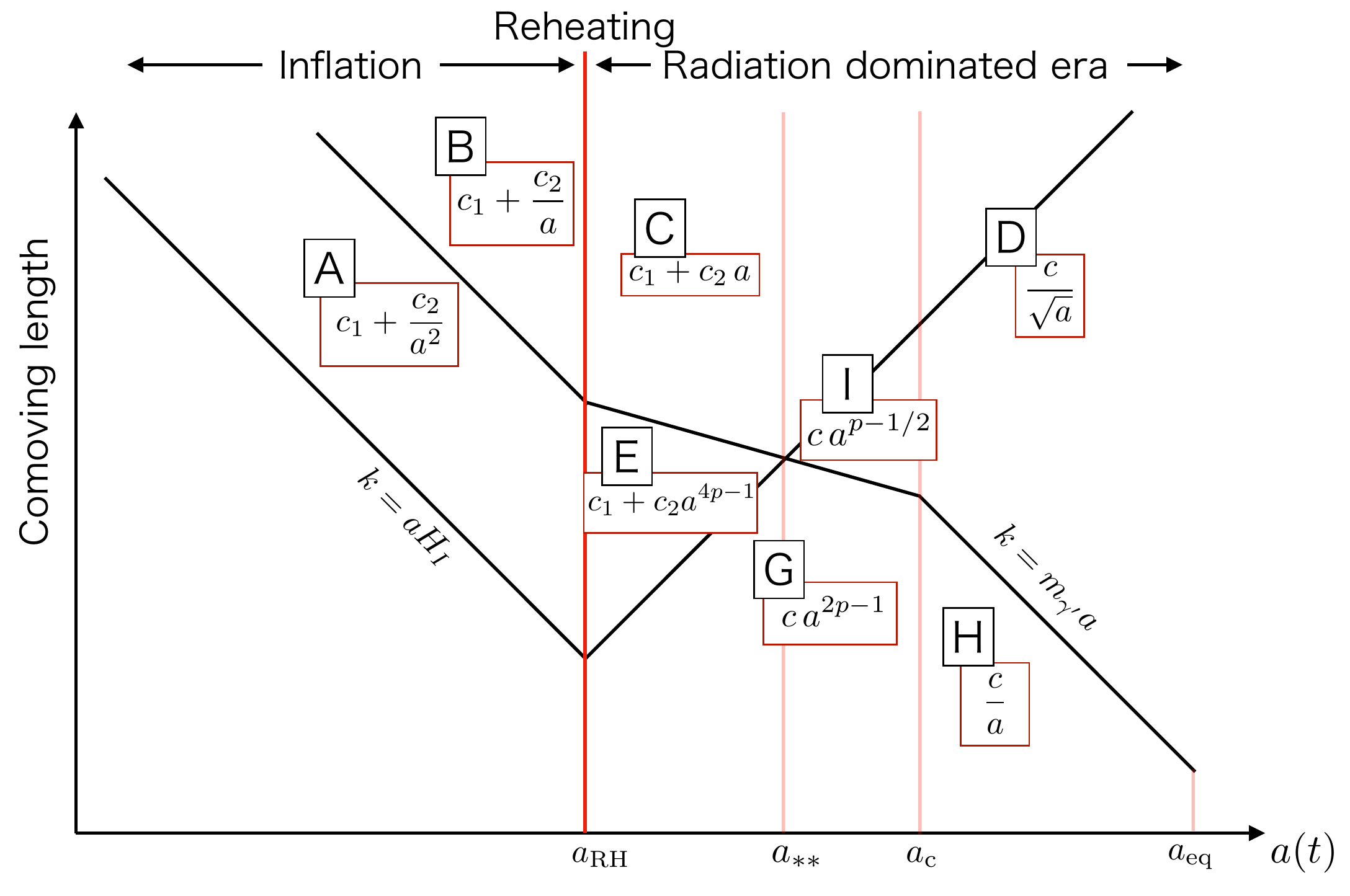}
\end{center}
\caption{ Schematic picture of comoving length evolution for Case 2. 
\label{fig:1-2}
}
\end{figure}

Next, we solve the equation of motion in the super-horizon/sub-horizon and relativistic/non-relativistic regime. 
The uppercase alphabets (A,B,C,$\dots$,I) represent different regimes. They are also presented in the lower panel of Fig.~\ref{fig:1}. 

\begin{itemize}

\item \textbf{Super-horizon relativistic regime (A, E, F)} - $H\gg k/a\gg \mg$, \\
During inflation, the relativistic (massless) mode 
obeys the Bunch-Davis vacuum solution \eq{eom for AL}. 
After the horizon exit, the solution is approximately given by 
\begin{align}
A_{L}(\vec{k},t)
\simeq 
\Aex(\vec{k})\lkk1+\frac{k^2}{2a^2H_{I}^2}
\rkk \,,
\label{initialA_L}
\end{align}
or simply 
\begin{equation}
 A_L \propto a^0, \ a^{-2} \,, 
 \quad \text{(regime A)} \,,
\end{equation}
for $k \ll a \HI$.

During the radiation dominated era, we obtain $2 (\partial_t m_{\gamma'})/\mg =- 4p H(t)$ for $a(t) < \ac$ 
and $0$ for $a(t) > \ac$. Then, the equation of motion is reduced to be
\begin{align}
&\lkk \partial_{t}^2+\lmk 3-4p \rmk H\partial_{t} \rkk A_{L} \simeq 0
&\text{for} \ a(t) < \ac 
\\
&(\partial_{t}^2 + 3H\partial_{t})A_{L} \simeq 0
&\text{for} \ a(t) > \ac  \,.
\end{align}
The solutions to these equations are given by 
\begin{align}
&A_{L} \propto a^0, \ a^{4p-1}
&\text{for} \ a(t) < \ac 
 \quad \text{(regime E)}
\\
&A_{L} \propto a^0, \ a^{-1}
&\text{for} \ a(t) > \ac 
 \quad \text{(regime F)} \,.
\end{align}
Because the super-horizon mode is approximately constant during inflation, the mode stays constant during the radiation-dominated era.

\item \textbf{Sub-horizon relativistic regime (G, H)} - $k/a\gg \mg,H$\\
In this regime, we can neglect $m_{\gamma'}$ (except for the one in the friction term) and take $H \del_t$ as a small parameter. 
The equation of motion is expressed as 
\begin{align}
&\lkk \partial_{t}^2+ \lmk 3-4p \rmk H\partial_{t}+\frac{k^2}{a^2} \rkk A_L \simeq 0 
&\text{for} \ a(t) < \ac 
\\
&\lmk \partial_{t}^2+3 H\partial_{t}+\frac{k^2}{a^2} \rmk A_L \simeq 0 
&\text{for} \ a(t) > \ac  \,,
\end{align}
or equivalently,
\begin{align}
&\lkk \partial_{\eta}^2+ \lmk 2-4p \rmk a H\partial_{\eta}+k^2 \rkk A_L \simeq 0 
&\text{for} \ a(t) < \ac 
\\
&\lmk \partial_{\eta}^2+2 a H\partial_{\eta}+k^2 \rmk
A_L \simeq 0 
&\text{for} \ a(t) > \ac  \,,
\end{align}
where we define the conformal time $\eta$ by $d \eta = d t/a$. 
The solutions to these equations are approximately given by 
\begin{align}
\label{SupRD}
&A_{L} \propto a^{2p-1} e^{ik\eta}, \ a^{2p-1} e^{-ik\eta}
&\text{for} \ a(t) < \ac 
 \quad \text{(regime G)}
\\
&A_{L} \propto a^{-1} e^{ik\eta}, \ a^{-1} e^{-ik\eta}, 
&\text{for} \ a(t) > \ac 
 \quad \text{(regime H)} \,.
\end{align}

\item\textbf{Late-time non-relativistic regime (D, I)} - $\mg\gg k/a,H$ \\
In this regime, the $k^2$ term in the friction term can be neglected, and we obtain 
\begin{equation}
(\partial_{t}^2+H\partial_{t}+\mg^2(t) )A_{L} \simeq 0 \,.
\end{equation}
The solution to this equation is approximately given by 
\begin{align}
&A_{L}\propto a^{p-1/2} \exp\lkk \pm \frac{i \mginf t }{1-p} \lmk \frac{t_{\rm RH}}{t} \rmk^{1-p} \rkk, 
&\text{for} \ a(t) < \ac 
 \quad \text{(regime I)} \,.
 \\
 &A_{L}\propto a^{-1/2} e^{i\mg t}, \ a^{-1/2} e^{-i\mg t}
&\text{for} \ a(t) > \ac 
 \quad \text{(regime D)} \,.
\end{align}

\item\textbf{Hubble-damped, non-relativistic regime (B, C)} - $H\gg \mg \gg k/a$\\
In this regime, we can neglect the $k^2/a^2$ and $m_{\gamma'}^2$ terms. The equation of motion is expressed as: 
\begin{equation}
(\partial_{t}^2+H\partial_{t})A_{L} \simeq 0 \,.
\end{equation}
The solution to this equation is given by 
\begin{equation}
A_{L} \propto a^0, \ a^{-1},
 \quad \text{(regime B)} \,,
\end{equation}
during inflation and 
\begin{equation}
A_{L} \propto a^0, \ a^{1}, 
 \quad \text{(regime C)} \,,
\end{equation}
during the radiation-dominated era.

\end{itemize}

\subsection{Summary of the spectrum for longitudinal mode (Case 1)}

The scale-factor dependence of $A_L(k,t)$ for each $k$ is presented in the lower panel of Fig.~\ref{fig:1} for Case 1. 
The constants $c_i$ and $c$ should be understood as different values in different regimes and determined by connecting the solutions on the boundaries. 
In the super-horizon regime, 
the solution stays constant for $a < a_*$ (regime A, B, C, E, F), 
and then decreases as $\propto a^{-1/2}$ for $a > a_*$ (regime D). 
In the sub-horizon regime, 
the solution evolves as $\propto a^{2p-1}$ for $\aRH < a < \ac$ (regime G), decreases as $\propto a^{-1}$ for $\ac < a < \aNRl$ (regime H), and decreases as $\propto a^{-1/2}$ for $a > \aNRl$ (regime D). 
Hence, the solution behaves similar to the case with a constant $\mg$, except for the regime $G$, which is the case for $k > \ac H(\ac) = \kRH (\m/\mginf)^{1/(2p)}$. 
This difference leads to another peak at a small scale as we will see shortly.

\begin{figure}[t]
\begin{center}
\includegraphics[clip, width=11cm]{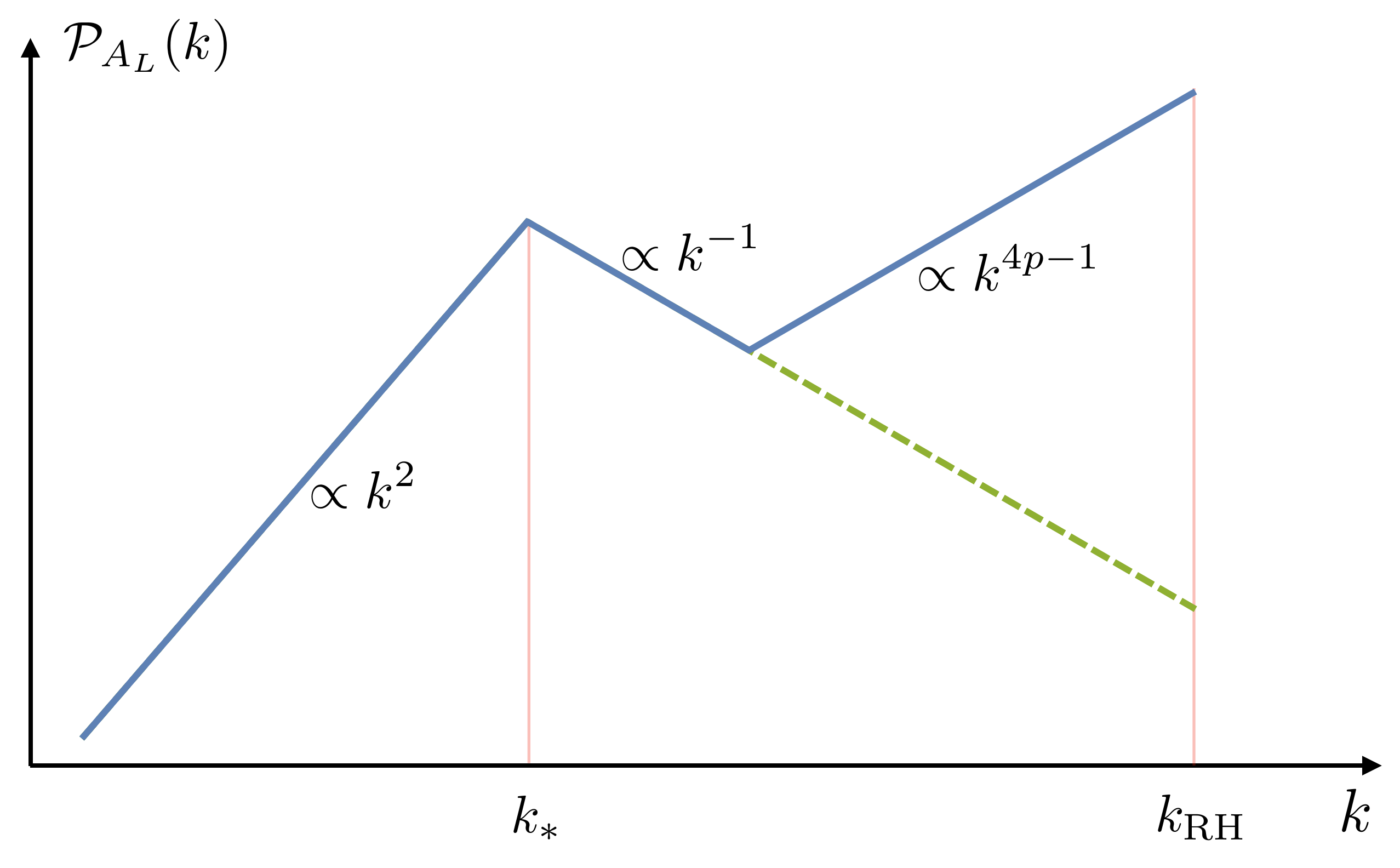}
\end{center}
\caption{ Schematic diagram of the spectrum of $A_L$ (blue line) for Case 1.
There are two peaks at $k = \kstar$ and at $k = \kRH$. 
The dashed line represents the case for the St\"{u}ckelberg mechanism considered in Ref.~\cite{Graham:2015rva}. 
\label{fig:spectrum}
}
\end{figure}

From the above results, we calculate the value of $A_L$ at the matter-radiation equality $a = \anow$. 
We can divide the wavenumber into three regimes: \begin{align}
&{\rm Regime \, a}: \quad k < \kstar
\\
&{\rm Regime \, b}: \quad \kstar < k < \kRH \lmk \frac{\m}{\mginf} \rmk^{1/(2p)}
\\
&{\rm Regime \, c}: \quad \kRH \lmk \frac{\m}{\mginf} \rmk^{1/(2p)} < k < \kRH \,. 
\end{align}
Here, we note that the mode with $k > \kRH$ is exponentially suppressed because it is always sub-horizon throughout the history of the Universe. 
Assuming that 
the mode $\kRH$ is already non-relativistic by $a = \anow$, 
we obtain 
\begin{align}
&{\rm Regime \, a}: \quad 
\frac{A_L(\anow)/\anow}{\Aex/\aex} = 
\lmk \frac{\aex}{\anow} \rmk 
\lmk \frac{\anow}{\astar} \rmk^{-1/2} 
\\
&\qquad \qquad \qquad \qquad \qquad \quad ~= 
\lmk \frac{k}{\kRH} \rmk 
\lmk \frac{\HI}{\m} \rmk^{1/4}
\lmk \frac{\Hnow}{\HI} \rmk^{3/4}
\\
&{\rm Regime \, b}: \quad 
\frac{A_L(\anow)/\anow}{\Aex/\aex} = 
\lmk \frac{\aex}{\anow} \rmk 
\lmk \frac{\aNRl}{\ain} \rmk^{-1} 
\lmk \frac{\anow}{\aNRl} \rmk^{-1/2} 
\\
&\qquad \qquad \qquad \qquad \qquad \quad ~= 
\lmk \frac{k}{\kRH} \rmk^{-1/2} 
\lmk \frac{\m}{\HI} \rmk^{1/2}
\lmk \frac{\Hnow}{\HI} \rmk^{3/4}
\\
&{\rm Regime \, c}: \quad 
\frac{A_L(\anow)/\anow}{\Aex/\aex} = 
\lmk \frac{\aex}{\anow} \rmk 
\lmk \frac{\ac}{\ain} \rmk^{2p-1} 
\lmk \frac{\aNRl}{\ac} \rmk^{-1} 
\lmk \frac{\anow}{\aNRl} \rmk^{-1/2} 
\\
&\qquad \qquad \qquad \qquad \qquad \quad ~= 
\lmk \frac{k}{\kRH} \rmk^{2p-1/2} 
\lmk \frac{\mginf}{\m} \rmk
\lmk \frac{\m}{\HI} \rmk^{1/2}
\lmk \frac{\Hnow}{\HI} \rmk^{3/4} \,, 
\end{align}
where we use $\aRH / a_{\rm eq} = (\HI / H_{\rm eq})^{1/2}$. 
The spectrum of the longitudinal mode is therefore given by the blue line in Fig.~\ref{fig:spectrum}, where the dashed line represents the one for the case with a constant $\mg$. 
Because the spectrum at a large scale is suppressed by $\propto k^2$, the isocurvature constraint from the CMB observations is circumvented. 
We have peaks at $k = \kstar$ and $\kRH$ for $p >1/4$: 
\begin{align}
\label{eq:peak1}
 &\left. \frac{A_L(\anow)/\anow}{\Aex/\aex} \right\vert_{k = \kstar} = \lmk \frac{\m}{\HI} \rmk^{1/4}
\lmk \frac{\Hnow}{\HI} \rmk^{3/4}
 \\
\label{eq:peak2}
 &\left. \frac{A_L(\anow)/\anow}{\Aex/\aex} \right\vert_{k = \kRH} = 
 \lmk \frac{\mginf}{\m} \rmk
\lmk \frac{\m}{\HI} \rmk^{1/2}
\lmk \frac{\Hnow}{\HI} \rmk^{3/4}\,.
\end{align}
The latter one is the maximum for $(\mginf/\m) > (\HI/\m)^{1/4}$.%
\footnote{
\label{footnote1}
It is stated that several e-folding numbers are necessary for super-horizon modes to be classicalized after the horizon exit (see, e.g.,Ref.~\cite{Barvinsky:1998cq}). This implies that the mode that enters the horizon soon after the end of inflation may not be classicalized. This introduces an uncertainty for the peak wavenumber and corresponding amplitude of \eq{eq:peak2}. We omit this factor for simplicity and assume that all modes are classicalized soon after they exit the horizon. 
}

\subsection{Summary of the spectrum for longitudinal mode (Case 2)}

Next, we calculate the value of $A_L$ at the matter-radiation equality $a = \anow$ for Case 2. 
In this case, we have the regime I instead of the regime F (see Fig.~\ref{fig:1-2}) and a mode with a small wavenumber grows as $\propto a^{p-1/2}$ from $a = a_{**}$ to $a = a_c$. 
We can divide the wavenumber into three regimes: 
\begin{align}
&{\rm Regime \, a}: \quad k < \kstars
\\
&{\rm Regime \, b}: \quad \kstars < k < \kRH \lmk \frac{\m}{\HI} \rmk 
\lmk \frac{\m}{\mginf} \rmk^{-1/(2p)}, 
\\
&{\rm Regime \, c}: \quad 
\kRH \lmk \frac{\m}{\HI} \rmk 
\lmk \frac{\m}{\mginf} \rmk^{-1/(2p)}
< k < \kRH \,. 
\end{align}
Assuming that 
the mode $\kRH$ is already non-relativistic by $a = \anow$, 
we obtain 
\begin{align}
&{\rm Regime \, a}: \quad 
\frac{A_L(\anow)/\anow}{\Aex/\aex} = 
\lmk \frac{\aex}{\anow} \rmk 
\lmk \frac{\ac}{\astars} \rmk^{p-1/2}
\lmk \frac{\anow}{\ac} \rmk^{-1/2} 
\\
&\qquad \qquad \qquad \qquad \qquad \quad ~= 
\lmk \frac{k}{\kRH} \rmk 
\lmk \frac{\HI}{\mginf} \rmk^{(1-2p)/(4-4p)}
\lmk \frac{\mginf}{\m} \rmk^{1/2}
\lmk \frac{\Hnow}{\HI} \rmk^{3/4}
\\
&{\rm Regime \, b}: \quad 
\frac{A_L(\anow)/\anow}{\Aex/\aex} = 
\lmk \frac{\aex}{\anow} \rmk 
\lmk \frac{\aNRm}{\ain} \rmk^{2p-1} 
\lmk \frac{\ac}{\aNRm} \rmk^{p-1/2} 
\lmk \frac{\anow}{\ac} \rmk^{-1/2} 
\\
&\qquad \qquad \qquad \qquad \qquad \quad ~= 
\lmk \frac{k}{\kRH} \rmk^{2p-1/2} 
\lmk \frac{\mginf}{\m} \rmk
\lmk \frac{\m}{\HI} \rmk^{1/2}
\lmk \frac{\Hnow}{\HI} \rmk^{3/4}
\\
&{\rm Regime \, c}: \quad 
\frac{A_L(\anow)/\anow}{\Aex/\aex} = 
\lmk \frac{\aex}{\anow} \rmk 
\lmk \frac{\ac}{\ain} \rmk^{2p-1} 
\lmk \frac{\aNRl}{\ac} \rmk^{-1} 
\lmk \frac{\anow}{\aNRl} \rmk^{-1/2} 
\\
&\qquad \qquad \qquad \qquad \qquad \quad ~= 
\lmk \frac{k}{\kRH} \rmk^{2p-1/2} 
\lmk \frac{\mginf}{\m} \rmk
\lmk \frac{\m}{\HI} \rmk^{1/2}
\lmk \frac{\Hnow}{\HI} \rmk^{3/4} \,.
\end{align}
Regimes b and c have the same wavenumber dependence. 
Accordingly, we have a peak at $k = k_{**}$ for $p<1/4$ and at $\kRH$ for $p > 1/4$: 
\begin{align}
\label{eq:peak3}
 &\left. \frac{A_L(\anow)/\anow}{\Aex/\aex} \right\vert_{k = \kstar} = \lmk \frac{\m}{\HI} \rmk^{1/4} 
\lmk \frac{\Hnow}{\HI} \rmk^{3/4}
 \lmk \frac{\mginf}{\mginf^{\rm (th)}} \rmk^{(3/4)/(1-p)}
 \\
\label{eq:peak4}
 &\left. \frac{A_L(\anow)/\anow}{\Aex/\aex} \right\vert_{k = \kRH} = 
 \lmk \frac{\mginf}{\m} \rmk
\lmk \frac{\m}{\HI} \rmk^{1/2}
\lmk \frac{\Hnow}{\HI} \rmk^{3/4}\,,
\end{align}
where we use \eq{eq:mginfth1} for $\mginf^{\rm (th)}$. 
Note that $\mginf > \mginf^{\rm (th)}$ in Case 2 (see \eq{eq:mginfth2}).

\section{Dark photon DM}
\label{sec:DM}

\subsection{Dark photon abundance}

From Eqs.~(\ref{energy density spectrum2}) and (\ref{spectrum1}), 
we can calculate the energy density of dark photons at $a = \anow$, which is expressed as 
\begin{equation}
 \rho_{\gamma'} \simeq 
 \frac{\HI^4}{(2\pi)^2} 
 \lmk \frac{\m}{\mginf} \rmk^2 
 \lmk \frac{A_L(\anow)/\anow}{\Aex/\aex} \rmk^2_{k=\kstar, \kRH} \,.
\end{equation}
Note that the dark photon abundance is suppressed by the mass squared difference during inflation and at present, owing to the rescaling of the canonical field for the longitudinal modes, \eq{rescale}. 
Substituting Eqs.~(\ref{eq:peak1}), (\ref{eq:peak2}), (\ref{eq:peak3}), and (\ref{eq:peak4}) into this, we obtain 
\begin{align}
 &\frac{\Omega_{\gamma'}h^2}{\Omega_{\rm DM}h^2} 
 \sim R \lmk \frac{\m}{\mginf} \rmk^2 
 \lmk \frac{\m}{6 \times 10^{-6} \eV} \rmk^{1/2} 
 \lmk \frac{\HI}{10^{14} \GeV} \rmk^2
\quad &\text{for} \ 
\frac{\mginf}{\m} < \lmk \frac{\HI}{\m} \rmk^{1/4}
 \label{result1}
\\
 &\frac{\Omega_{\gamma'}h^2}{\Omega_{\rm DM}h^2} 
 \sim 
 \lmk \frac{\m}{0.8 \GeV} \rmk 
 \lmk \frac{\HI}{10^{14} \GeV} \rmk^{3/2} 
 \quad &\text{for} \ 
\frac{\mginf}{\m} > \lmk \frac{\HI}{\m} \rmk^{1/4}\,, 
 \label{result2}
\end{align}
where $R$ is defined shortly, $\Omega_{\rm DM}$ is the observed DM density parameter, and $h$ ($\simeq 0.67$) is the Hubble parameter at present in the unit of $100\, {\rm km/sec/Mpc}$. We take $\Omega_{\rm DM} h^2 \simeq 0.12$. 
Here we note that our result has $\mathcal{O}(1)$ uncertainty because of the approximation we made. In Ref.~\cite{Graham:2015rva}, they calculated the dark photon abundance numerically for the case with a constant $\mg$ and found a numerical correction to the analytic calculations. To make our result consistent with their results, we multiply $\rho_{\gamma'}$ by a factor of about $0.18$ in the above results. 
We define 
\begin{align}
 R &\equiv 
 {\rm Max} \lkk 1, \ \lmk \frac{\mginf}{\mginf^{\rm (th)}} \rmk^{(3/2)/(1-p)} \rkk\,,
 \\
 &= {\rm Max} \lkk 1, \ 
 \lmk \frac{\HI}{\m} \rmk^{3/2} \g^{(3/2)/(1-p)}  \rkk\,,
\end{align}
where we use $\mginf = \g \HI$ 
and \eq{eq:mginfth1} for $\mginf^{\rm (th)}$. 
These are the main result of this paper. 
We note that the dark photon mass should become constant by the time of matter-radiation equality. 
This puts an upper bound on $\mginf$ such as 
\beq
 a_c \ll a_{\rm eq}
 \quad 
 \leftrightarrow 
 \quad 
 \g \ll \frac{\m}{H_{\rm eq}^p \HI^{1-p}}\, ,
 \label{eq:upperbound}
\eeq
where we use $\mginf = \g \HI$. 
The Hubble parameter at the matter-radiation equality is given by $H_{\rm eq} \simeq 2.26 \times 10^{-37} \GeV$.

Taking $\Omega_{\gamma'} h^2= \Omega_{\rm DM}h^2$, 
we obtain the following conditions to explain the DM abundance using the gravitationally produced dark photon:
\begin{align}
\label{result3}
 &\m \simeq 0.8 \GeV \times 
 {\rm Min} \lkk R^{-2/5} \lmk \frac{\g}{3 \times 10^{-11}} \rmk^{4/5}, \ 
 \lmk \frac{\HI}{10^{14} \GeV} \rmk^{-3/2} 
 \rkk \,,
\end{align}
where we use $\mginf = \g \HI$.
The former (latter) one originates from the peak at $k = k_*$ ($k_{\rm RH}$). 
We note that the latter one may contain a theoretical uncertainty of $\mathcal{O}(1\,\text{-}\,10)$ from the process of classicalization for super-horizon modes (see footnote~\ref{footnote1}).

\begin{figure}[t]
\begin{center}
\includegraphics[clip, width=11cm]{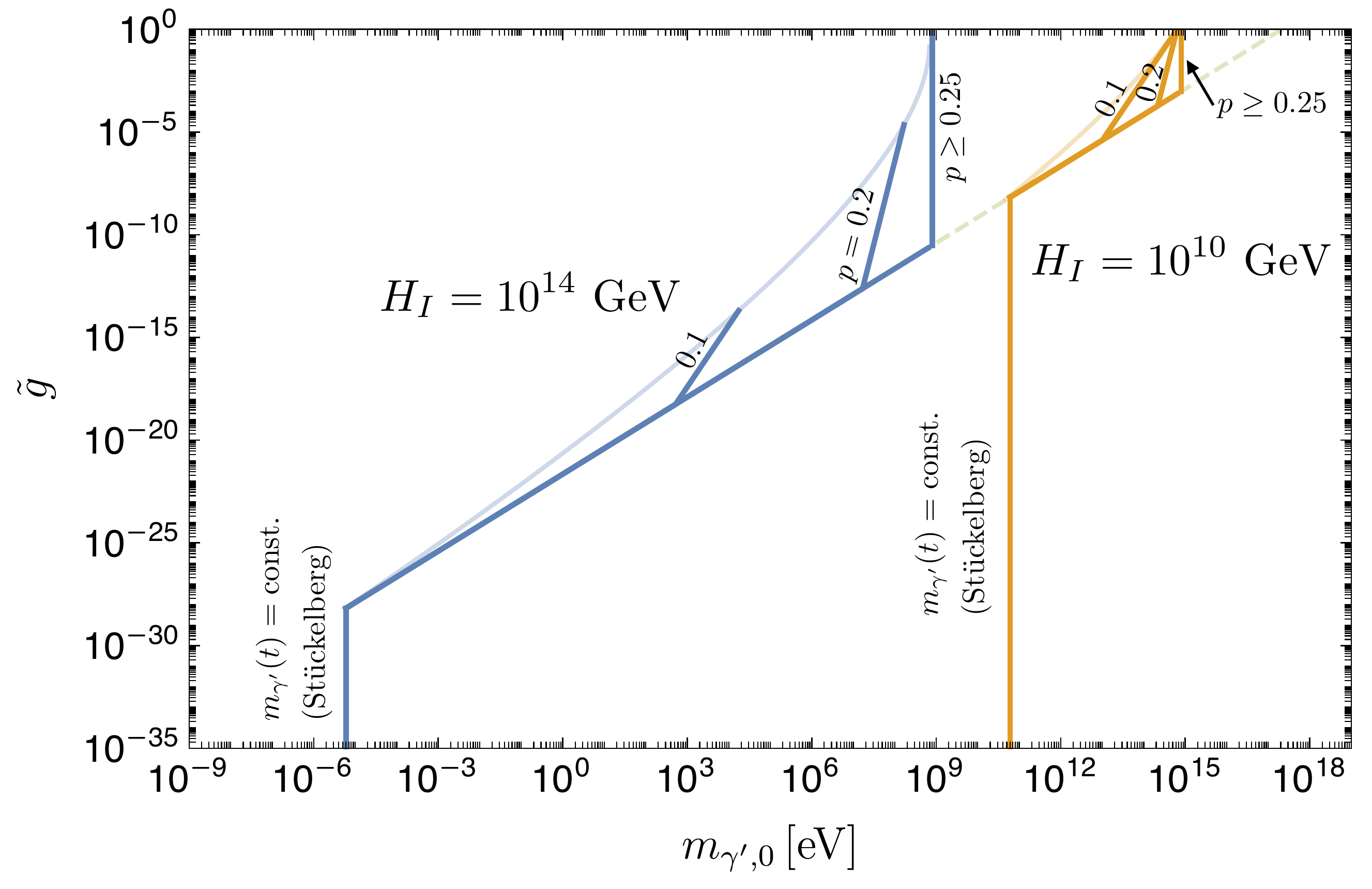}
\end{center}
\caption{Modified gauge coupling constant $\g$ ($\lesssim g$) as a function of $\m$ to explain the DM abundance. We assume $\HI = 10^{10}$ and $10^{14} \GeV$ for the orange and blue lines, respectively. 
We plot the cases with $p = 0.1$, $0.2$, and $0.25$, where 
the case with $0.25 < p \le 1$ is the same as the one with $p = 0.25$. 
The thin solid curves represents an upper bound of \eq{eq:upperbound} for a given $p$. 
The thin dashed line represents the contribution solely from the peak at $k = k_*$ with $R=1$. 
The region with a larger $\m$ (to the right of the lines) is excluded because of the gravitational overproduction of dark photon. 
\label{fig:summary}
}
\end{figure}

We plot the required value of $\g$ as a function of $\m$ for $H_I = 10^{10} \GeV$ (orange) and $10^{14}\GeV$ (blue) in Fig.~\ref{fig:summary}. 
We plot the cases with $p = 0.1$, $0.2$, and $0.25$, where 
the case with $0.25 < p \le 1$ is the same as the one with $p = 0.25$. 
For $p < 0.25$, the dependence is given by 
the first term in \eq{result3} with $R =1$ and $R > 1$ for an intermediate and large $\g$, respectively. 
In this case, $\g$ has an upper bound from the condition \eq{eq:upperbound}, which is represented by the thin solid curves. 
For $p \ge 0.25$, 
the dependence is given by 
the first and the second terms in \eq{result3} for an intermediate and large $\g$, respectively. 
In this case, $\g$ can be as large as $\mathcal{O}(1)$. 
For a negligible $\g$ (see \eq{conditionStuckelberg}), 
the dark photon mass is always constant and the result is given by the one obtained for the St\"{u}ckelberg mechanism case, as will be discussed shortly. 
The thin dashed line represents the one solely from the first term in \eq{result3} with $R=1$, which presents the absolute upper bound of $\m$ as $2 \times 10^8 \GeV$ for $\g =\mathcal{O}(1)$.

Finally, 
we check whether $\kstar$ and $\kstars$ are horizon-in by the time of matter-radiation equality. 
This is the case when 
\begin{align}
&\kstar \gtrsim a_{\rm eq} H_{\rm eq}
 ~~ \leftrightarrow ~~ 
 \m \gtrsim H_{\rm eq} \,, 
 \\
 &\kstars \gtrsim a_{\rm eq} H_{\rm eq}
 ~~ \leftrightarrow ~~ 
 \g \gtrsim \lmk \frac{H_{\rm eq}}{\HI} \rmk^{1-p} \,. 
\end{align}
This is satisfied in the most parameter space of interest. 
We should also check if the mode of $\kRH$ becomes non-relativistic by the time of matter-radiation equality. 
This is the case when 
\begin{equation}
 \kRH \lesssim \m a_{\rm eq}
 ~~ \leftrightarrow ~~ 1 \lesssim \frac{\m^2}{\HI H_{\rm eq}} 
 \sim 
 3\times 10^{22} \lmk \frac{\m}{0.8 \GeV} \rmk^2 
 \lmk \frac{\HI}{10^{14} \GeV} \rmk^{-1} \,. 
\label{condition-kstar}
\end{equation}
This is satisfied when the mode at $k = \kRH$ dominates and $\Omega_{\gamma'} = \Omega_{\rm DM}$. Hence, \eq{result3} is justified. 
If \eq{condition-kstar} is not satisfied, 
the mode $k = \kRH$ behaves like dark radiation. Its energy density is approximately given by $\Omega_{\gamma'}/\Omega_{\rm rad} \sim H_I^4 / (H_I^2 \Mpl^2)$ and 
is significantly smaller than the radiation in the Universe.

\subsection{Relation to the case with St\"{u}ckelberg mechanism}
\label{sec:Stuckelberg}

There is a lower bound on $\g$, to justify our calculation above. 
We assumed $\m < \mginf$ because, otherwise, the Higgs stays at the low-energy minimum during inflation, and this scenario is essentially identical to the case with the St\"{u}ckelberg mechanism. 
The condition implies that 
\begin{equation}
\label{constraint1}
 \g \gtrsim \frac{\m}{\HI} \,. 
\end{equation}
Together with \eq{result3}, 
we obtain 
\begin{align}
 &\m \gtrsim 6 \, \mu {\rm eV} 
 \lmk \frac{\HI}{10^{14} \GeV} \rmk^{-4} \,. 
 \label{result5}
\end{align}
If this condition is not satisfied, the abundance is given by the one calculated in Ref.~\cite{Graham:2015rva}: 
\begin{equation}
\label{result-previous}
\frac{\Omega_{\gamma'}h^2}{\Omega_{\rm{DM}}h^2}
\sim \lmk \frac{\m}{6 \, \mu{\rm eV}} \rmk^{1/2}
\lmk\frac{H_{I}}{10^{14}\GeV}\rmk^2 \,,
\end{equation}
because the dark photon mass is constant in the entire history of the Universe. 
Note that the threshold of \eq{result5} is simply given by the one to explain the DM abundance for the case of the St\"{u}ckelberg mechanism. 
This is represented by the vertical lines at a small $\g$ in Fig.~\ref{fig:summary}. 
Hence, we conclude that the dark photon mass should be larger than or equal to $6 \, \mu {\rm eV}$, to explain the DM abundance. Compared with the case of the St\"{u}ckelberg mechanism, the dark photon mass is not necessarily related to the inflation energy scale. Therefore, the parameter space is broadened for the Higgs mechanism case.

We obtain an upper bound on the dark photon mass and a lower bound on the Hubble parameter during inflation from Eqs.~(\ref{result3}) and (\ref{result5}). Combining them, we obtain 
\begin{align}
 &\m \lesssim 2 \times 10^8 \GeV \,,
 \\
 &H_I \gtrsim 2 \times 10^8 \GeV \,,
\end{align}
where we use $R \ge 1$. 
If these conditions are not satisfied, the dark photon abundance is determined by \eq{result-previous}. This can be observed from \eq{constraint1}, where the condition cannot be satisfied for $\g \lesssim \mathcal{O}(1)$, for such a small $H_I$.

Using \eq{result-previous} with $\Omega_{\gamma'} h^2 = \Omega_{\rm DM} h^2$, 
we obtain 
\begin{equation}
\label{conditionStuckelberg}
 \g \lesssim 6 \times 10^{-29} \lmk \frac{\HI}{10^{14} \GeV} \rmk^{-5} \,,
\end{equation}
for the case of the constant dark photon mass. 
Hence, the negligible gauge coupling constant and/or a relatively small inflation energy scale are/is required to realize a constant dark photon mass with the Higgs mechanism throughout the history of the Universe. 
This parameter space is represented as vertical lines at a small $\g$ in Fig.~\ref{fig:summary}.

\subsection{Consistency with weak gravity conjecture}

The very small gauge coupling constant may not be compatible with the weak gravity conjecture~\cite{Arkani-Hamed:2006emk}, which is a consistency condition for a low-energy effective field theory with quantum gravity. 
According to Ref.~\cite{Reece:2018zvv}, 
the conjecture requires the following condition: 
\beq
 H_I \lesssim g^{1/3} M_{\rm Pl}, 
\eeq
where $M_{\rm Pl}$ is the Planck mass. 
We note that $g \lesssim \g$ as we discussed below \eq{eq:mass}, so that we obtain 
\beq
 \g \gtrsim 6 \times 10^{-16} \lmk \frac{H_I}{10^{14} \GeV} \rmk^3. 
 \label{eq:WGC}
\eeq
This condition can be satisfied for $\m \gtrsim 0.1 \MeV$ for $R =1$ even if $H_I$ is as large as $10^{14} \GeV$.

If one adopts the model of \eq{Vphi}, 
one should also require that $\Lambda$ is larger than $H_I$ and is smaller than $g^{1/3} \Mpl$. 
This condition is stronger than the above one if $\Lambda$ is much larger than $H_I$.

We note that the condition \eq{eq:WGC} is not satisfied 
in the case of the St\"{u}ckelberg mechanism, as it may correspond to a smaller gauge coupling constant of \eq{conditionStuckelberg}~\cite{Reece:2018zvv,Draper:2022pvk}. 
In fact, the scenario with the St\"{u}ckelberg mechanism is not consistent with the weak gravity conjecture unless $H_I \lesssim 3 \times 10^{12} \GeV$ or $\mg \gtrsim 60 \eV$. 
On the contrary, the scenario with the Higgs mechanism can work for a relatively large gauge coupling constant and can be consistent with the weak gravity conjecture even for $H_I = 10^{14} \GeV$.

\subsection{Observational constraints and implications}

The dark photon generically has a kinetic mixing with the U(1)$_Y$ gauge boson. 
In the presence of bi-charged particles, one expects that the kinetic mixing arises from a one-loop effect expressed as (see, e.g., Ref.~\cite{Gherghetta:2019coi})
\begin{equation}
 \epsilon = - \frac{g_Y g}{16 \pi^2} 
 \sum_i Y_i q_i \ln \frac{M_i^2}{\mu^2} \,, 
\end{equation}
where $g_Y$ denotes the U(1)$_Y$ gauge coupling, 
$Y_i$ ($q_i$) represent U(1)$_Y$ (U(1)$'$) charges of fields in the loop with mass $M_i$, 
and $\mu$ is a renormalization scale. This leads to $\epsilon \sim (10^{-2}\, \text{-}\, 10^{-1}) \times g$. 
Since the gauge coupling must be very small to explain DM in our scenario (see Fig.~\ref{fig:summary} and note $g \lesssim \tilde{g}$),
such kinetic mixing is far below the experimental sensitivity in the mass region below $1$\,MeV.
If a larger kinetic mixing is caused by some mechanism, dark photon DM could be discovered in future experiments (see Ref.~\cite{Caputo:2021eaa}).

A dark photon a with mass larger than approximately $1\MeV$ 
decays into an electron-positron pair through the kinetic mixing. 
The kinetic mixing parameter must be strongly suppressed for such a dark photon to be DM. This can be realized if the theory satisfies a hidden $Z_2$ parity under which  the dark photon and the dark Higgs flip the sign. If this $Z_2$ parity is weakly broken, we expect a small kinetic mixing. The small kinetic mixing results in the decay of the dark photon into the SM particles and provides some indirect detection signals of DM~\cite{Chen:2008yi,Chen:2008qs,Chen:2008md}.

\section{Discussion and conclusions}
\label{sec:conclusion}

In this paper we have investigated the gravitational production of dark photons by quantum fluctuations during inflation. 
In particular, we have considered the case in which the dark photon obtains a mass via the Higgs mechanism, that is, the spontaneous symmetry breaking of the U(1)$'$ gauge symmetry, which is never restored after inflation. 
To realize this scenario, the mass of the Higgs field should be higher than  the Hubble parameter during inflation, $H_I$. 
Then, due to the unitarity of the Higgs coupling, the Higgs VEV should also be larger than $H_I$. If the Higgs boson is so heavy that it is effectively decoupled during and after inflation, the dark photon mass is constant
throughout the history of the Universe. This is equivalent  to the St\"uckelberg mass
considered in Ref.~\cite{Graham:2015rva}, and in this case,
the dark photon mass should satisfy $6 \, \mu {\rm eV}$ for $H_I = 10^{14} \GeV$ to explain DM~\cite{Graham:2015rva}. 
This implies a large hierarchy between $H_I$ and the dark photon mass, or equivalently, a very small gauge coupling. 
Instead, we have focused on the case in which the Higgs boson acquires a tachyonic Hubble-induced mass, and the Higgs VEV changes after inflation such that it is larger than $H_I$ but is much smaller at present. 
In this case, the dark photon has a time-dependent mass after inflation, and its time evolution drastically changes compared to the case of a time-independent mass. We have analytically derived the scale-factor dependence of the amplitude of the longitudinal mode in each regime of interest.

Assuming instantaneous reheating, we have estimated the spectrum of the longitudinal mode of the dark photons. It shows peaks at an intermediate scale and a small scale, at the wavenumber that enters the horizon when the dark photon becomes non-relativistic, and at the wavenumber that enters the horizon just after the end of inflation. 
The dark photon abundance is dominated by the former wavenumber for a small gauge coupling and inflation scale, whereas it can be dominated by the latter wavenumber for a relatively large gauge coupling and inflation scale. 
If the inflation scale is as large as $10^{14} \GeV$, the dark photon mass should be equal to or smaller than $0.8 \GeV$ in order to explain the DM abundance. 
For an extremely small gauge coupling, 
the dark photon mass is independent of time and 
our result is reduced to the case of the St\"{u}ckelberg mechanism. This yields the lower bound of the dark photon mass. In this sense, our scenario is a natural extension for the case with the St\"{u}ckelberg mechanism 
and provides a larger parameter space for the dark photon DM. 

There are two significant differences between the results of this study and that of the case with the St\"{u}ckelberg mechanism. 
First, the abundance of the longitudinal modes of dark photons is suppressed by the ratio between the dark photon mass squared during inflation and that at present. This is because the longitudinal modes can be represented by a scalar field by rescaling the field using a mass and its wavenumber in the relativistic regime. Therefore, gravitational production is relatively suppressed by the mass of the dark photons during inflation, which 
requires a larger amplitude of dark photon and its gauge coupling to explain the DM abundance. 
This effect is relatively reduced if its mass continues to change even after it becomes non-relativistic. 
The second difference is 
the presence of the second peak at a small scale, which 
can be understood from the picture of the NG boson that is absorbed by the longitudinal mode. In the sub-horizon relativistic regime, the NG boson (i.e., the phase direction of the Higgs field) has a nonzero rotational velocity in the phase space. When the radial direction of the Higgs field decrease over time, 
the rotational velocity of the NG boson increase because of the conservation of angular momentum in the phase space~\cite{Kobayashi:2016qld}. This results in the 
relative enhancement of the longitudinal modes of the dark photon, which implies that a mode with a larger wavenumber grows over a longer time, so that the resulting spectrum has a peak at a large wavenumber.

We have found that the dark-photon spectrum may have another peak at the small scale, depending on the time evolution of the dark-photon mass. 
It is the mode that enters the horizon just after the end of inflation. Such a large wavenumber mode may or may not have enough time to be cliassicalized during inflation. If we observe the dark photon with such a mode, we may be able to obtain some information for the mechanism of classicalization of quantum fluctuations during inflation. This is a unique way to reveal physics at the end of inflation via the observation of DM.

{\it Note added: 
After completing our paper, a paper~\cite{Gorghetto:2022sue} appeared on arXiv, where they discussed 
the formation of dark photon stars (vector solitons)~\cite{Jain:2021pnk,Amin:2022pzv} and their implication for vector DM substructure 
for the case in which the vector DM with a constant mass is generated gravitationally. The same conclusion should apply to our case at least for a small gauge coupling regime because our calculation reduces to the case with the St\"{u}ckelberg mechanism in that limit. 
}

\section*{Acknowledgments}
This work was supported by MEXT Leading Initiative for Excellent Young Researchers (M.Y.), and by JSPS KAKENHI Grant No.\ 20H01894 (F.T.), 20H05851 (F.T. and M.Y.), and 21K13910 (M.Y.), and JSPS Core-to-Core Program (grant number: JPJSCCA20200002) (F.T.).

\appendix

\section{Case with $p=1$}
\label{sec:appendix}

In this Appendix, we particularly consider the case with $p=1$ and discuss that the resulting spectrum for the longitudinal mode is given by the one derived in the main text by substituting $p = 1$ and $R =1$. 

We first note that the case with $p = 1$ corresponds to the case with $n = 2$ in the model of \eq{Vphi}. 
In this case, the Higgs field may not follow the potential minimum and the symmetry is restored because its perturbations grow to overcome the potential barrier at the origin of the potential. 
One may consider the case with a rapid decay of Higgs field that introduces a friction term on the Higgs equation of motion. In this case, the Higgs would follow its potential minimum. However, the backreaction from the produced particles leads to a large thermal mass term of the Higgs field, which exceeds the Hubble induced mass term and makes U(1)$'$ symmetry restored at a later time. 
These issues are present even in the case of $n < 5$, which corresponds to $p > 0.25$. 
We still consider the case even in this parameter space as there may be other models that are consistent with such a parameter space.

\begin{figure}[t]
\begin{center}
\includegraphics[clip, width=14cm]{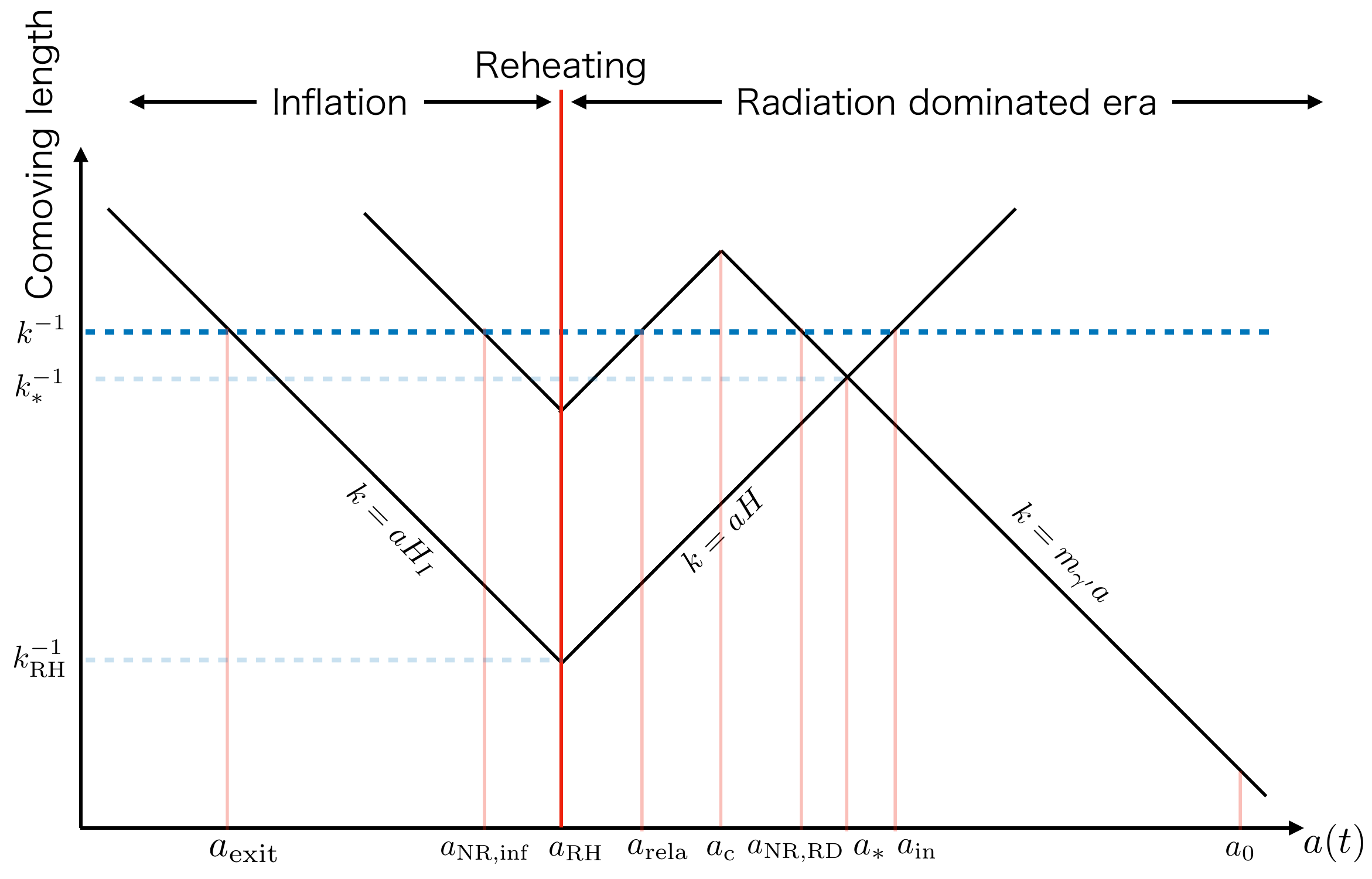}
\end{center}
\vspace{0.3cm}
\begin{center}
\includegraphics[clip, width=14cm]{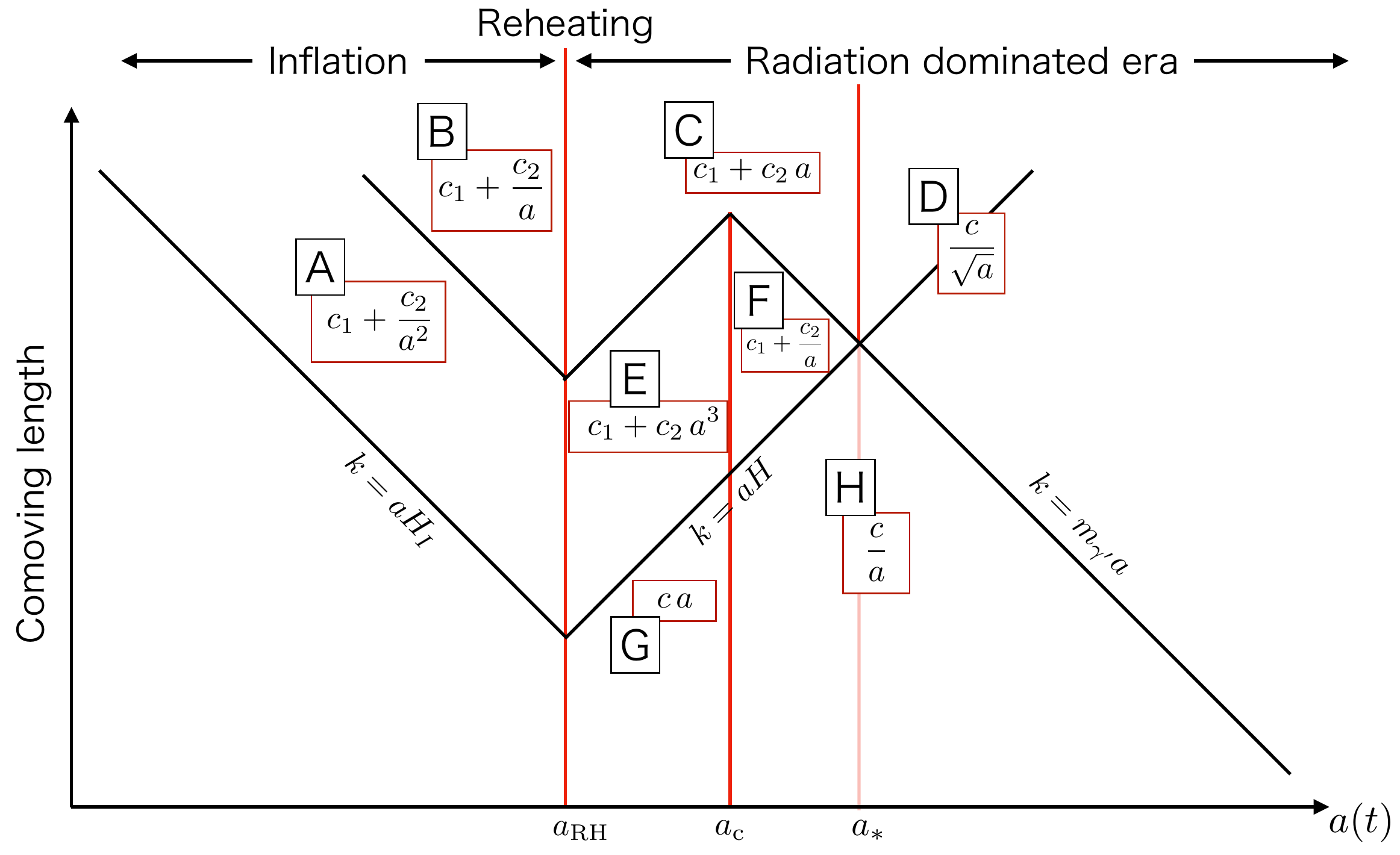}
\end{center}
\caption{Same as Fig.~\ref{fig:1} but with $p=1$. In this case, a mode can become non-relativistic twice. 
\label{fig:2}
}
\end{figure}

As we did in the main text, we divide into the regimes in which we can analytically solve the equation of motion. 
The threshold of the relativistic/non-relativistic limit is complicated owing to the time-dependence of the dark photon mass. 
A mode with a fixed comoving wavenumber $k$ can become non-relativistic during inflation at 
$a(t) = \aNRi \equiv (\HI/\mginf) (k/\kRH) \aRH$ because its physical wavenumber $k/a(t)$ decreases in time while the dark photon mass is constant. 
After inflation, the dark photon mass decreases as $\propto t^{-1} \propto a(t)^{-2}$, which is faster than $a(t)^{-1}$, such that the mode can become relativistic at $a(t) = \arela \equiv (\mginf/\HI) (\kRH/k) \aRH$. 
At $a(t) = \ac$, the dark photon mass becomes constant. 
Subsequently, the physical wavenumber decreases faster than the dark photon mass; hence, the mode can become non-relativistic at $a = \aNR \equiv (k/\kRH)(\HI/\m) \aRH$. 
This dependence is summarized in the upper panel of Fig.~\ref{fig:2}, where the mode on the blue-dashed line becomes non-relativistic twice. The vertical red line represents the reheating epoch at $a(t) = \aRH$. 
Although the meaning of the scale factors are qualitatively different, 
they have the same parameter dependence as Eqs.~(\ref{eq:scale1}), (\ref{eq:scale2}), and (\ref{eq:scale3}) with the replacement of $\aNRe \to \aNRi$, $\aNRm \to \arela$, and $\aNRl \to \aNR$ with $p=1$. 
This implies that the calculations and results in the main text can also be applied even for $p=1$.

The scale-factor dependence of $A_L(k,t)$ for each $k$ is presented in the lower panel of Fig.~\ref{fig:2}. 
The behavior of $A_L(k,t)$ is similar to the case for $p < 1/2$ 
and the resulting spectrum is given by Fig.~\ref{fig:spectrum} with $p =1$. 
Although there may be a correction to the spectrum 
from the initially subdominant term, which decreases as $A_L \propto a^{-2}$ after the horizon exit during inflation, 
we have checked that the peak amplitudes do not change by the contribution from the initially subdominat term. 
The resulting dark photon abundance is therefore given by Eqs.~(\ref{result1}) and (\ref{result2}) with $R=1$.

\bibliography{references}

\end{document}